# Timing Model Extraction for Sequential Circuits Considering Process Variations

Bing Li, Ning Chen, Ulf Schlichtmann
Institute for Electronic Design Automation, Technische Universitaet Muenchen, Germany
{b.li, ning.chen, ulf.schlichtmann}@tum.de

## ABSTRACT

As semiconductor devices continue to scale down, process variations become more relevant for circuit design. Facing such variations, statistical static timing analysis is introduced to model variations more accurately so that the pessimism in traditional worst case timing analysis is reduced. Because all delays are modeled using correlated random variables, most statistical timing methods are much slower than corner based timing analysis. To speed up statistical timing analysis, we propose a method to extract timing models for flip-flop and latch based sequential circuits respectively. When such a circuit is used as a module in a hierarchical design, the timing model instead of the original circuit is used for timing analysis. The extracted timing models are much smaller than the original circuits. Experiments show that using extracted timing models accelerates timing verification by orders of magnitude compared to previous approaches using flat netlists directly. Accuracy is maintained, however, with the mean and standard deviation of the clock period both showing usually less than 1% error compared to Monte Carlo simulation on a number of benchmark circuits.

## 1. INTRODUCTION

As the feature size of semiconductor device scales to deep submicron region, relative parameter variations increase. For example, the ratio of $3\sigma$ to nominal value of gate length increases from 34.6% at the 130nm node to 47.1% at the 70nm node, as shown in [13]. As a result of the increasing variations, traditional worst case static timing analysis (STA) becomes too pessimistic, where all parameters are set to their worst case values to evaluate circuit performance. Therefore, statistical static timing analysis (SSTA) is introduced to model process variations and analyze circuit performance more accurately. In SSTA, cell delays are modeled as functions of random variables which represent process parameters with variations. Then, arrival times are propagated to compute the circuit delay. Unlike the result of STA, the circuit delay in SSTA is a distribution providing delay-yield information, with which designers have a chance to make trade-off between performance and yield of the circuit.

As an emerging methodology, many SSTA algorithms have been introduced in past years. Assuming that cell delays are linear functions of Gaussian random variables, first-order methods were proposed in [1, 7, 18]. This linear assumption simplifies arrival time propagation algorithms at the expense of accuracy. To improve modeling and propagation accuracy, the canonical linear form in [18] was extended in [2] to handle non-Gaussian parameters and nonlinear delay functions. With the same purpose, quadratic methods were proposed in [5, 8, 19, 20]. Another method to improve timing accuracy was proposed in [16], where delays are modeled as linear functions of Gaussian and non-Gaussian random variables. The latter are identified by independent component analysis. In addition to the SSTA methods above, algorithms for analyzing timing performance of latch based circuits were introduced in [3, 21]. Because of latch transparency and feedback loops, these algorithms are much slower.

Although existing SSTA methods can generate more useful timing results, they exhibit runtime problem. In SSTA, all cell delays and arrival times are random variables and correlated to each other. This makes the addition and maximum/minimum computation in arrival time propagation complex and slow. As a speedup method, incremental timing analysis is used in [18], where the cells which are not in the output cone of revised cells are not visited. But this method still needs to visit all the cells in the output cone, whose number may still be large when the revised cells are close to the inputs of the circuit. To solve this problem, the circuit can be treated as hierarchical as it is in most cases. Consider a module is in the output cone of the revised cells. From the circuit view, the propagation of arrival times inside this module is unimportant, if the arrival times at the outputs of this module are correct and the timing requirements inside this module can be met. These timing requirements include setup time and hold time constraints for sequential cells. From this observation, a timing model can be extracted to replace this module for timing verification when there is no cell revised inside it. The timing model contains only the constraints of timing requirements inside this module and the information to compute arrival times at its outputs. Therefore, the timing model can be much smaller compared to the original netlist, because the internal circuit structure is not considered. Another advantage of using timing models is that it can benefit the designs using IP (Intellectual Property) macros from third-party vendors, where the complete netlists of IP macros are not always available because of IP protection. Instead, timing models can be provided as replacement for hierarchical timing analysis.

Considering timing model extraction, there are already many methods proposed for STA, where process variations are not taken into account. For combinational circuits, methods were proposed in [9, 12, 22] to transform a netlist to a much smaller one by discarding structural details, but maintaining the same input-output delays. For sequential circuits, the method in [12] extracts timing models by delay arc and check arc merge operations. To allow arbitrary level of latch transparency, all latch input pins are retained. The drawback of this method is that the extracted timing models are still latch based, so that a complex latch timing analysis algorithm, like the one proposed in [15], is still needed when such timing models are used. Another method to extract timing models for latch based circuits was proposed in [17], where timing constraints at the inputs of a module are abstracted. Because latches can be transparent, this method substitutes the constraints iteratively across latches. To reduce complexity, latch transparency level is assumed to be a predefined value. This assumption is too strict because latch transparency can not be fixed at design time. Instead, only the setup and hold time constraints of a latch based circuit guarantee its proper behavior. Because of process variations, the transparency of latches can even be different between chips after manufacturing, as shown in [6], but the functions of these chips are still correct. This is a remarkable advantage of latch based circuits when variations are taken into account. Another work to extract timing models for latch based circuits is [4], where a graph arrangement method is used to determine the timing constraints at inputs of a module.

When variations are considered, most of the timing model extraction methods for STA can not work properly. For combinational circuits, the delay patterns needed by the methods in [9, 22] do not exist, because delays are represented by correlated random variables instead of fixed values. For sequential circuits, the methods in [4, 17] depend on the clock period being a known fixed value, so that are not feasible for SSTA. As a solution, timing model extraction for combinational circuits considering process variations was proposed in [10]. Additionally, the correlation between modules is handled by a variable replacement method in [10]. The limitation of [10] is that it does not address sequential circuits.

The main contribution of this paper is that we propose a timing model extraction method for sequential circuits considering process variations. The timing model extraction for flip-flop based circuits is relatively simple and will be briefly introduced. The main part of this paper is timing model extraction for latch based circuits, where no assumption about the transparency level from an input is made. The extracted timing models are very small compared with the original netlists and preserve very good accuracy for hierarchical timing analysis. With such timing models, the runtime of hierarchical timing verification can be drastically reduced.

The rest of the paper is organized as follows. In Section 2 we will introduce the concept of a reduced timing graph and explain the method to extract timing models for flip-flop based circuits. In Section 3, timing model extraction for latch based circuits is proposed. This section is the main part of this paper. Experimental results by applying the proposed method to ISCAS89 benchmark circuits are shown in Section 4. Finally, we conclude our work in Section 5.

## 2. TIMING MODEL EXTRACTION FOR FLIP-FLOP BASED CIRCUITS

In this section, we will explain the concept of a reduced timing graph, which we will then use to introduce our method. Thereafter, timing model extraction for flip-flop based circuits will be explained. We will only consider setup time constraints in the following sections for simplicity, but our method can be easily adapted to include hold time constraints too. We will also omit the delays of direct paths between primary inputs and outputs. These delays can be handled e.g. by the techniques proposed in [10].

A sequential circuit is composed of sequential cells, e.g. flip-flops or latches, and combinational cells, which form the delay paths between sequential cells. In a *reduced timing graph* [21], a node represents a sequential cell, or a primary input, or a primary output. An edge represents the maximum delay between a pair of sequential cells, or the maximum delay from a primary input to a sequential cell, or from a sequential cell to a primary output. When process variations are considered, all edge delays

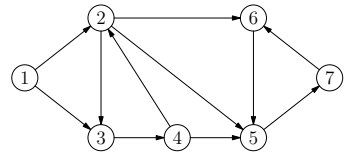

**Figure 1: Reduced timing graph example**

are random variables. Figure 1 shows a reduced timing graph as an example.

In order to explain our method, we define the following notations.

$\Delta_{ij}$ : edge delay between node $i$ and $j$ in a reduced timing graph.

$q_i$ : propagation delay of a sequential cell.

$s_i$ : setup time of a sequential cell.

$T$ : clock period.

Because of process variations, $\Delta_{ij}$, $q_i$ and $s_i$ are all random variables. In SSTA, the target is to compute the yield of a circuit at different clock periods. Therefore, we can not make any assumption about $T$ except that it is an unknown fixed value.

In the following, we will review timing constraints for a flip-flop based circuit. From these constraints, we can extract very simple timing models to accelerate timing verification at higher level. For two flip-flops $i$ and $j$ in the reduced timing graph, if there is a direct edge from $i$ to $j$, we call $i$ a fanin flip-flop of $j$, and $j$ is a fanout flip-flop of $i$. At the launching clock edge, the output of $i$ is updated to the value at its input with delay $q_i$. Thereafter, this value propagates to the input of $j$ with delay $\Delta_{ij}$. In order to guarantee the flip-flop $j$ to work properly, the latest data at its input must be stable at $s_j$ time before its launching clock edge, i.e.,

$$\max_i \{q_i + \Delta_{ij}\} \leq T - s_j \iff \max_i \{q_i + \Delta_{ij} + s_j\} \leq T \quad (1)$$

where the maximum is performed with all the fanin flip-flops $i$ of $j$ in the reduced timing graph. At any flip-flop in a circuit, the setup time constraint like (1) should be met. Therefore, the setup time constraint for a circuit can be written as

$$C_F : \quad \max_{i,j} \{q_i + \Delta_{ij} + s_j\} \leq T \iff D_F \leq T \quad (2)$$

where the maximum is performed with all flip-flop pairs with a direct edge in the reduced timing graph.

In addition to flip-flop fanins, a flip-flop may have primary inputs as fanins. Similar to the timing constraint for flip-flop pairs, the arrival time from a primary input $k$ to a flip-flop $j$ should meet

$$\hat{a}_k + \Delta_{kj} \leq T - s_j \iff \hat{a}_k + \Delta_{kj} + s_j \leq T \quad (3)$$

where $\hat{a}_k$ is the latest arrival time at primary input $k$ relative to the current clock period of $j$ and can be determined only when the circuit is integrated into a design as a module. When extracting the timing model for a circuit, no assumption should be made about $\hat{a}_k$. For primary input $k$, there may be more than one fanout flip-flop. The arrival times at the inputs of all these fanout flip-flops must meet the constraint (3), so that we have the timing constraint for a primary input $k$ as

$$C_{I_k} : \quad \max_j \{\hat{a}_k + \Delta_{kj} + s_j\} \leq T \iff \quad (4)$$

$$\hat{a}_k + \max_j \{\Delta_{kj} + s_j\} \leq T \iff \quad (5)$$

$$\hat{a}_k + D_{I_k} \leq T \quad (6)$$

where $D_{I_k}$ is computed with all the flip-flops $j$ which have direct edges coming from the primary input $k$ in the reduced timing graph.

When a circuit with $m$ primary inputs is used as a module in a hierarchical design, the probability that the complete circuit works can be computed as

$$Y = Prob\{C_a, C_F, C_{I_1}, \ldots, C_{I_m}\} \quad (7)$$

where $C_a$ represents the timing constraint for the other modules in the design, and can also be described by the timing constraints from their timing models. The probability in (7) is computed with all the constraints $C_a, C_F, C_{I_1}, \ldots, C_{I_m}$ are true at the same time.

In a hierarchical design, if a module is not changed during design iteration, we need not always to verify the timing constraints of all the flip-flops inside this module. Instead, the circuit yield can be evaluated using (7), where the constraints $C_F$ and $C_{I_1}, \ldots, C_{I_m}$ can be verified very fast when using $D_F$ and $D_{I_1}, \ldots, D_{I_m}$ directly. The verification acceleration comes from the fact that we do not need to re-compute $D_F$ and $D_{I_1}, \ldots, D_{I_m}$ for a module, if this module is not revised during design iteration. In our method, we compute these variables and provide them as the timing model for hierarchical timing verification.

The maximum delay $D_F$ can be computed effectively using a standard block-based SSTA algorithm by propagating arrival times from all flip-flops at the same time. For a primary input $k$, $D_{I_k}$ can be computed by propagating arrival times from $k$, where the arrival time $\hat{a}_k$ is temporarily set to 0. Finally, each constraint of $C_F$ and $C_{I_1}, \ldots, C_{I_m}$ is represented by a random variable respectively. When verifying the timing of a module, only these $m+1$ variables should be involved, which is much simpler than propagating arrival times across the original netlist.

When a circuit is used as a module in a hierarchical design, its outputs will be connected to the inputs of other modules. For example, when the primary output $v$ of a module is connected to the primary input $k$ of another module, the arrival time at $k$ is determined by the latest arrival time at $v$. In order to verify the timing constraints for the fanout flip-flops of $k$, a timing model should also contain the information of the data stable time at all its primary outputs. Normally the primary output $v$ has more than one fanin flip-flop. After the launching clock edge, data signals are propagated from all these fanin flip-flops $i$ to $v$. The data stable time or latest arrival time $D_{O_v}$ is then computed as

$$D_{O_v} = \max_i \{q_i + \Delta_{iv}\} \quad (8)$$

Assuming there are $n$ primary outputs in the module, the $n$ arrival times $D_{O_1}, \ldots, D_{O_n}$ should also be included in the timing model. Combining with the setup time constraints, the timing model for a flip-flop based circuit contains only $m+n+1$ random variables.

## 3. TIMING MODEL EXTRACTION FOR LATCH BASED CIRCUITS

In this section, we will explain how to extract timing constraints from latch based circuits. Using these constraints, the time-consuming timing verification algorithms in [3, 21] can be avoided.

### 3.1 Latch timing analysis

Because arrival times can propagate through latches, timing analysis of latch based circuits is more complex than that of flip-flop based circuits. When a data signal reaches the input of a latch during the active period of its clock, this data signal can be propagated through this latch instantly. This property is called latch transparency. With such property, a data signal can start to propagate to the next stage immediately when it reaches a transparent latch, thus allowing the delay of a combinational path to be larger than the clock period. This is called time borrowing or cycle stealing. A detailed example of this can be found in [11]. Like flip-flop, each latch $j$ also has setup time constraint $s_j$. Because of latch transparency an arrival time can propagate through many latch stages. At each stage, the arrival time must meet the setup time constraint of the corresponding latch. This is the source of the complexity of timing analysis for latch based circuits.

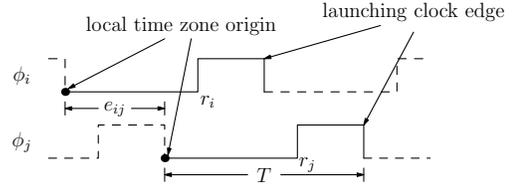

**Figure 2: Local time zone and clock phase shift**

To evaluate the timing performance of a latch based circuit, the complete timing constraints allowing multiphase clocks with the same period are specified in [15]. In this section, we will give a short review of these timing constraints and use them to explain our timing model extraction method.

In timing analysis for latch based circuits, all arrival times are represented in the *local time zone* [11], i.e., with respect to the starting time of the local time zone. In this paper, we assume the active clock level of latches is '1', and the starting time of each local time zone is the time when the clock signal of the latch switches from '1' to '0'. For two latches $i$ and $j$, where $i$ is fanin of $j$, their clock signals are illustrated in Figure 2. $e_{ij}$ is the *phase shift* of the two clocks, and will be used to transform an arrival time from the local time zone of $i$ to the local time zone of $j$.

Unlike flip-flop, where a data signal starts to propagate right after the launching clock edge, the time that a data signal starts to propagate to the next latch stage can be at any time when the clock is active. We call this time *departure time* and denote it as $d_i$ for latch $i$. Like arrival time, $d_i$ is also defined in the local time zone of latch $i$, i.e., it uses the origin of the local time zone as reference time 0. If we consider only the signal propagation from $i$ to $j$, the arrival time $a_j$ can be computed as

$$a_j = d_i + q_i + \Delta_{ij} - e_{ij} = d_i + m_{ij} \quad (9)$$

where $-e_{ij}$ transforms $a_j$ to the local time zone of $j$. $m_{ij}$ is the *delay shift* from $i$ to $j$. Considering all fanin latches $i$ of $j$, the arrival time $a_j$ can be computed as

$$a_j = \max_i \{d_i + m_{ij}\} \quad (10)$$

As in [11], the rising clock edge is called *enabling clock edge*. For latch $i$ we denote the time of its enabling clock edge as $r_i$ in the local time zone, as shown in Figure 2. Because a data signal can start to propagate to the next latch stage only after the enabling clock edge, the departure time of $i$ can be written as

$$d_i = \max\{a_i, r_i\} \quad (11)$$

By substituting (11) into (10), we can eliminate the departure time from (10) as

$$a_j = \max_i \{\max\{a_i, r_i\} + m_{ij}\} \quad (12)$$

In (12) all the fanins of latch $j$ are assumed to be latches. In fact $j$ may have primary inputs as fanins. With these primary inputs considered, the arrival time at the input of $j$ is extended to

$$a_j = \max\{\max_i\{\max\{a_i, r_i\} + m_{ij}\}, \max_k\{\hat{a}_k + \Delta_{kj}\}\} \quad (13)$$

where $\hat{a}_k$ is the arrival time at fanin primary input $k$ of $j$, and is represented in the local time zone of $j$. Similar to timing model extraction for flip-flop based circuits, we can make no assumption about $\hat{a}_k$.

The timing constraint for a latch $j$ is that the data signal at its input must be stable at least $s_j$ time before the launching clock edge. From this we can write the timing constraint for latch $j$ as

$$a_j \leq T - s_j \iff a_j + s_j \leq T \tag{14}$$

## 3.2 Timing constraint restructuring for latch based circuits

When a latch based circuit is used as a module in a hierarchical design, the constraint (14) should be checked for each latch inside the module. In order to accelerate this timing verification, we will propose a method to extract only the timing constraints needed by higher level timing analysis. With these constraints, the internal latches of a module need not to be visited, so that the complex timing verification algorithms for latch based circuits can be avoided.

As the first step, we will restructure the timing constraint (14) for all latches to a different form which is equivalent to (14). Based on such constraints, we will explain our timing model extraction method later. Consider timing constraints for latch $j$, we substitute $a_j$ in (14) with (13). The new constraint is equivalent to (15)-(17).

$$\max_i \{a_i + m_{ij}\} + s_j \leq T \tag{15}$$

$$\max_i \{r_i + m_{ij}\} + s_j \leq T \tag{16}$$

$$\max_k \{\hat{a}_k + \Delta_{kj}\} + s_j \leq T \tag{17}$$

where the first two maximum operations are performed with all fanin latches $i$ of $j$. The last maximum is performed with all fanin primary inputs $k$ of latch $j$. For a fanin latch $i$, we can continue to substitute $a_i$ in (15) with the form of (13) and split the constraint after this substitution into three parts similarly as

$$\max_p \{a_p + m_{pi}\} + m_{ij} + s_j \leq T \tag{18}$$

$$\max_p \{r_p + m_{pi}\} + m_{ij} + s_j \leq T \tag{19}$$

$$\max_q \{\hat{a}_q + \Delta_{qi}\} + m_{ij} + s_j \leq T \tag{20}$$

where the first two maximum operations are performed with all fanin latches $p$ of $i$, and the last maximum is performed with all fanin primary inputs $q$ of $i$.

From (18)-(20), we can observe that after each substitution, the arrival time in the setup time constraint is shifted by one latch stage backwards. Combining with (15), the constraint (19) defines that the data signals starting from the enabling clock edges of the latches two stages before $j$ should meet the setup time constraint at $j$, where all the latches in between are considered transparent. Similarly, the arrival time from any primary input in this latch stage range should also meet such setup time constraint, as defined by (20). By repeating the substitution backwards through all fanin latches recursively, we can find that the arrival times starting from the enabling clock edges of all latches in the fanin cone of $j$ must meet the setup time constraint of $j$. For any primary input in the fanin cone of $j$, we can infer similar constraints. Because each latch in the circuit has a constraint like (14), we can run the recursive substitution above for all the latches. Thereafter, if we switch the view point to an arrival time starting from the enabling clock edge of a latch, we can find that this arrival time must meet the setup time constraints of all latches in the fanout cone of the latch. From this observation, we can describe the timing constraints for arrival times from enabling clock edges of all latches together.

$L_1$: The arrival time from the enabling clock edge of any latch to all latches in the fanout cone of the latch must meet the setup time constraints of these latches, with all intermediate latches assumed transparent.

Similarly we can describe the timing constraints for primary inputs.

$L_2$: The arrival time from any primary input must meet the setup time constraints of all latches in the fanout cone of the primary input, with all intermediate latches assumed transparent.

If the substitution from latch $j$ is fulfilled across a loop through $j$ in the reduced timing graph with sufficient iterations, $j$ will eventually be reached again. In this case, the timing constraint becomes to

$$a_j + M_{j \to j} + s_j \leq T \tag{21}$$

where $M_{j \to j}$ is the sum of all the delay shifts across the loop in the reduced timing graph, called *cumulative delay shift*. Because $a_j$ in (21) can be substituted further, $M_{j \to j}$ must be nonpositive. Otherwise, the left side of (21) will exceed any $T$ after sufficient traversals of the loop. We call a loop in the reduced timing graph whose cumulative delay shift is nonpositive a nonpositive loop. From the discussion above, the third constraint for a latch circuit can be described as

$L_3$: All loops in the reduced timing graph must be nonpositive.

After a loop is traversed, the arrival time $a_j$ can be substituted further. But the timing constraints for enabling clock edges and primary inputs extracted in these further substitutions need not to be verified. As an example, the arrival time from the enabling edge of latch $i$ to latch $j$ after traversing a loop is smaller than the arrival time when $j$ is traversed the first time, so that the constraint after a loop is always dominated by the constraint before a loop is traversed, as shown in (22).

$$r_i + m_{ij} + M_{j \to j} + s_j \leq r_i + m_{ij} + s_j \leq T \tag{22}$$

With $L_3$ as condition, we can revise $L_1$ and $L_2$ to $L_{R1}$ and $L_{R2}$.

$L_{R1}, L_{R2}$: The constraints of $L_1$ and $L_2$ without visiting latches after loops, respectively.

After each substitution, we still get constraints from the arrival times of the fanin latches, like (18). If this substitution continues infinitely, the arrival times just after chip reset will be reached. These arrival times start from the corresponding enabling clock edges, as implicitly used in [14, 21]. This shows that the constraint set $L_{R1}$, $L_{R2}$ and $L_3$ can guarantee the setup time constraint (14) for all latches to be met. Because $L_{R1}$, $L_{R2}$, and $L_3$ are extracted from (14), this discussion proves that the constraint set $L_{R1}$, $L_{R2}$ and $L_3$ is equivalent to (14). Based on the results above, the yield of a hierarchical design using a latch based module can be written as

$$Y = Prob\{L_a, L_{R1}, L_{R2}, L_3\} \tag{23}$$

where $L_a$ is the timing constraint set for the latches in the other modules.

In the following, we will explain how to extract timing constraints needed by higher level timing verification for a latch based module. The basic idea is that, we replace each constraint $L_{R1}$, $L_{R2}$, $L_3$, with a simple form to avoid the complexity of verifying latch timing. As an example, we will compute the maximum loop cumulative delay shift and use it to represent $L_3$. During higher level timing verification, these loops need not to be enumerated again. Instead, only the provided variable needs to be verified against the clock period.

### 3.3 Timing constraint extraction from enabling clock edges

The constraint $L_{R1}$ defines that the arrival time starting from the enabling clock edge of any latch must meet the setup time constraints of all latches in the fanout cone of the latch without through loops, with all latches in between assumed as transparent. For simplicity, the arrival times we mention henceforth are all with the latch transparency assumption.

From the reduced timing graph in Figure 1, we can see that there are many paths starting from a latch. According to $L_{R1}$, the arrival times should be propagated through all these paths. In large circuits, using a direct path enumeration method is prohibitive. To solve this problem, a feedback loop breaking algorithm with heuristics is used in [3]. The basic idea of the feedback loop breaking is explained in the following. The reduced graph is searched in depth-first order. During this search, if some fanin edges of the current visited latch $i$ originate from latches which are in the fanout cone of $i$, these edges are removed from the reduced timing graph. The removed edges are called *feedback edges*, because there are loops through them starting from $i$ and ending at $i$. Consequently, the reduced timing graph changes to a directed acyclic graph. Arrival times can be propagated across this revised graph using a standard SSTA method. With latch 1 as starting latch, an example of the revised timing graph of Figure 1 is illustrated in Figure 3, where all nodes are assumed latches and feedback edges are shown with dashed arrows. Note that the result of feedback edge removal is not unique, depending on different traversal orders when searching feedback edges.

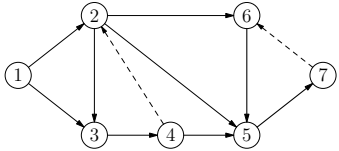

**Figure 3: Reduced timing graph example with feedback edge removal**

In Figure 3, the path 1→3→4→2→5→7→6 is missing when visiting latches from 1. To solve this problem, it is proposed in [3] to search the original reduced timing graph with different orders, so that the probability of missing paths can be reduced. In this paper, we simply run arrival time propagation twice to reduce the runtime of timing model extraction. In the second run, the latch visiting order is the same as in the first run. The arrival times at source nodes of feedback edges created in the first run are updated to their sink nodes and propagated further. In this way, any path with one feedback edge is guaranteed to be traversed. Like explained in [3], there are still missing paths with more than one feedback edge. From our experiments, traversing reduced timing graph twice already shows good accuracy for timing model extraction. In a reduced timing graph, paths with more than one feedback edge are relatively longer than other paths. As to be explained later, the arrival time propagation for $L_{R1}$ and $L_{R2}$ will stop when the arrival time is smaller than the enabling clock edge of a latch. Therefore, long path traversals are not completely required. When creating the constraint for $L_3$, any edge in the reduced timing graph will be visited at least once. Additionally, the delay shifts along a path will compensate each other. As a result, long paths have less chance to affect the constraint for $L_3$. This explains why the accuracy is still acceptable when traversing the reduced timing graph only twice. For better path coverage, the algorithm in [3] can also be used to replace the simple traversal in this paper.

Till now we have discussed the latch traversal order for extracting timing constraints from enabling clock edges. Like standard SSTA methods, we have two operations during this arrival time computation: addition and maximum. But the difference in our computation is that the delay shift $m_{ij}$ from latch $i$ to $j$ will be added to the arrival time. From the definition in (9), $m_{ij}$ is equal to $q_i + \Delta_{ij} - e_{ij}$. $q_i$ and $\Delta_{ij}$ are known random variables, so that their sum can be computed. But $e_{ij}$ can not simply be treated as a known random variable. As shown in Figure 2, $e_{ij}$ defines the clock phase shift. If the clock phases are generated using an absolute delay based method, $e_{ij}$ can be safely assumed as a known random variable. Therefore, the arrival time update is the same as in standard SSTA methods. Our method explained below can be adapted to handle timing constraint extraction in this case easily. If the clock phases are generated so that the relative clock phase shift is fixed, i.e., $e_{ij}$ has a fixed ratio to the clock period, the update is quite different. In the following, we will discuss the second case only.

In the second clock scheme, the clock phase shift changes proportionally when the clock period changes, so that we can write $e_{ij}$ as

$$e_{ij} = c_{ij}T \quad (24)$$

where $c_{ij}$ is a positive constant. Similarly, we assume the time of the enabling clock edge in the local time zone has a fixed ratio to the clock period, i.e.,

$$r_i = c_i T \quad (25)$$

where $c_i$ is a positive constant smaller than 1.

To extract setup time constraints from the enabling clock edge of latch $i$, we firstly set its arrival time to $r_i$. The arrival time from $i$ to any following latch $j$ can be written as

$$a_j = r_i + M_{i \to j} = D_{ij} + C_{ij}T \quad (26)$$

where $M_{i \to j}$ is the cumulative delay shift across the path from $i$ to $j$. By substituting all the clock phase shifts in $M_{i \to j}$ and $r_i$ in (26) with (24) and (25), we can merge all the known random variables and the coefficients of $T$ respectively and denote the sums with $D_{ij}$ and $C_{ij}$, where $D_{ij}$ is a random variable and $C_{ij}$ is a fixed value. In the following, we call $D_{ij}$ and $C_{ij}$ *delay part* and *coefficient part* of (26) respectively.

When more than one arrival time from fanin latches are merged at latch $j$, their maximum is computed. Each such arrival time is in the same form of (26). Because these arrival times reach $j$ through different paths, the delay and the coefficient of $T$ may be different. As a result, the maximum can not simply be performed with all these arrival times like in SSTA. Instead, only when two arrival times have the same coefficient of $T$, can they be merged to one by computing the maximum of their delay parts. If two arrival times have different coefficients for $T$, we simply propagate them further at the same time. As a result, an arrival time in our propagation becomes to a set of items. Each item in this set is represented by a random variable and the coefficient for $T$, in the form of (26). Any time when a delay shift is added to such an arrival time, this delay shift is added to each item in the arrival time, where the delays and the coefficients are added separately. When the maximum of two such sets is computed, the items with identical coefficient are merged. The other items which can not be merged are simply appended to the new arrival time and propagated further.

With arrival times computed, we can now explain how to create setup time constraints for each latch. The maximum of all the items in an arrival time must meet the setup time constraint of the corresponding latch. For each item, the constraint can be written as

$$D_{ij} + C_{ij}T \leq T - s_j \iff \quad (27)$$
$$D_{ij} + s_j \leq (1 - C_{ij})T \iff \quad (28)$$
$$(D_{ij} + s_j)/(1 - C_{ij}) \leq T \quad (29)$$

where $1 - C_{ij}$ is positive because $C_{ij}$ is computed by subtracting the coefficient of $T$ when traversing each latch stage, as shown in (9).

In (29) all the random variables and coefficients on the left side are known, so that $(D_{ij} + s_j)/(1 - C_{ij})$ can be treated as a known random variable. At each latch during arrival time propagation, we create such a constraint in the form of (29) for each item in the arrival time. After propagating arrival times from enabling clock edges of all latches, all these constraints together form the constraint of $L_{R1}$. Because all the variables in these constraints should be smaller than $T$ to guarantee the correct circuit behavior with clock period $T$, these constraints together are equivalent to the constraint that the maximum of all the random variables at the left side of them is smaller than $T$. This maximum is denoted as $V_1$, and the constraint $L_{R1}$ can be simply written as

$$V_1 \leq T \quad (30)$$

During arrival time propagation, each arrival time is represented by a set of items in the form of (26). As the propagation recurs further, the number of items in arrival times will increase, whence decelerates the computation of $V_1$. In the following, we will explain how to reduce the number of items in an arrival time of a latch. Based on our discussion before, the arrival time from any enabling clock edge is propagated and setup time constraint is included implicitly in (30). During the propagation, if an item from an arrival time is smaller than the time of the enabling clock edge in that local time zone, the constraint created from propagating this item further is dominated by the constraint created from the arrival time propagation starting from the enabling clock edge. Therefore, we can remove such an item from the arrival time without affecting the timing constraint represented by (30). The condition for removing an item is described as following,

$$D_{ij} + C_{ij}T \leq r_j = c_j T \quad (31)$$

If $c_j - C_{ij}$ is positive, (31) is equivalent to

$$D_{ij}/(c_j - C_{ij}) \leq T \quad (32)$$

Because the left side of (32) is a random variable, the condition (32) can be true only with a certain probability. During arrival time propagation, $V_1$ increases gradually while the constraint (29) is merged to (30) at each latch. To merge a constraint, the maximum of $V_1$ and the random variable at the left side of (29) is computed. $V_1$ is then updated with the result. When verifying the timing performance of a circuit, the constraint (30) will be true. Comparing (32) with (30), we can remove the arrival time item if the following condition is met.

$$D_{ij}/(c_j - C_{ij}) \leq V_1 \quad (33)$$

Both sides of (33) are random variables, so that (33) can be true only with a certain probability. If the probability that (33) is true approximates 1, the removal of the corresponding arrival time item affects the timing model only with very small probability. We compute the probability in (34) for each arrival time item during propagation.

$$p_r = Prob\{D_{ij}/(c_j - C_{ij}) \leq V_1\} \quad (34)$$

If $p_r$ is larger than a predefined constant $\delta$ approximating 1, we remove the item from the arrival time.

Like the timing model extraction for flip-flop based circuits described in Section 2, the timing model for a latch based circuit should contain delays to the primary outputs. During the arrival time propagation in this section, if the fanout of a latch is a primary output, the delay from this latch to the primary output and the arrival time are added together and stored as the output delay. From the analysis in Section 3.2, we know that any arrival time can be considered as starting from an enabling clock edge or from a primary input initially. In the former case, if an arrival time item can reach an output without being removed at an intermediate latch, this arrival time item should be verified against the setup time constraint of the latches in the following modules. This explains our method to extract output delays from internal latches.

### 3.4 Timing constraint extraction from primary inputs

After the timing constraint representing $L_{R1}$ is explained in Section 3.3, we will explain the timing constraints from primary inputs, i.e., finding a simple form to represent $L_{R2}$.

The basic idea to extract timing constraint for a primary input is mostly the same as the one described in Section 3.3. The arrival time from a primary input is propagated across the reduced timing graph with feedback edge removal. At each latch, the maximum of the arrival times is computed and the setup time constraint is updated.

The only difference of this traversal from the one in Section 3.3 is that the starting arrival timing from the primary input $k$ is $\hat{a}_k$, which is unknown until this module is integrated into a hierarchical design. Consequently, an arrival time item from $k$ to a latch $j$ becomes to

$$a_j = \hat{a}_k + D_{kj} + C_{kj}T \quad (35)$$

If this arrival time is propagated a latch stage further, the corresponding delay shift can be merged with the right side of (35) by adding the random variables and the coefficients of $T$ separately. Because all arrival time items are in the form of (35) and share the same $\hat{a}_k$, the maximum of two of them can be performed just like the maximum computation in Section 3.3 without considering $\hat{a}_k$. The result of this maximum computation is still an item set with $\hat{a}_k$ implicitly appended.

At each latch, the setup time constraint from each item in the arrival time is extracted. An example of such a constraint is shown below.

$$\hat{a}_k + D_{kj} + C_{kj}T \leq T - s_j \iff \quad (36)$$
$$\hat{a}_k + (D_{kj} + s_j) + C_{kj}T \leq T \quad (37)$$

To represent $L_{R2}$, we create a constraint set $\mathcal{C}_k$ for the primary input $k$. Each item from such a constraint set is in the form of (37). Because $\hat{a}_k$ is unknown, we can not simplify (37) like from (28) to (29). Instead, we directly insert this constraint into the constraint set $\mathcal{C}_k$. During this insertion, if there is already a constraint item inside $\mathcal{C}_k$ with the same coefficient of $T$, we need only to merge the random variable $D_{kj} + s_j$ in (37) with the corresponding variable of the constraint item. Otherwise, a new constraint is simply inserted into $\mathcal{C}_k$.

Similar to compressing arrival times in Section 3.3, we compare each arrival time item with the time of the enabling clock edge. An example of such comparison for latch $q$ is shown in (38).

$$\hat{a}_k + D_{kq} + C_{kq}T \leq r_q = c_q T \quad (38)$$

Consider that there is already a set of constraints $\mathcal{C}_k$ for the primary input $k$. Each item in this set is in the form of (37). If

we subtract both sides of (37) from (38), we have

$$(D_{kq} - D_{kj} - s_j) + (C_{kq} - C_{kj})T \leq (c_q - 1)T \iff \quad (39)$$
$$D_{kq} - D_{kj} - s_j \leq (c_q - 1 - C_{kq} + C_{kj})T \quad (40)$$

If (39) is true, the arrival time item can be removed because (38) is dominated by (37). If $c_q - 1 - C_{kq} + C_{kj}$ is positive, (40) is equivalent to

$$(D_{kq} - D_{kj} - s_j)/(c_q - 1 - C_{kq} + C_{kj}) \leq T \quad (41)$$

Similar to (32)-(34), if the probability $p_i$ is larger than $\delta$, we can remove the arrival time item, where $p_i$ is defined as

$$p_i = Prob\{(D_{kq} - D_{kj} - s_j)/(c_q - 1 - C_{kq} + C_{kj}) \leq V_1\} \quad (42)$$

Like in Section 3.3, the output delays to primary outputs are also created if a fanout is a primary output during propagation. The only difference is that the output delays depend on the arrival time $\hat{a}_k$ at primary inputs.

### 3.5 Nonpositive loop constraint extraction

The last constraint for a timing model is $L_3$, which specifies all feedback loops in the reduced timing graph should be nonpositive. In this paper, we will adapt the two-run traversal method used in Section 3.3 to compute the maximum loop delays, although other loop breaking algorithms, e.g. [3], can also be used for better path coverage.

The basic idea is to compute the maximum arrival time starting from each latch and looping back to it again. At first, we set the arrival time at the starting latch to 0, and propagate arrival times using the two-run traversal in Section 3.3, but without updating latch setup time constraints. During the propagation, if a fanout latch is the starting latch, a loop is formed. In this case, we add the delay shift between the current latch and the starting latch to the arrival time of the current latch to compute the maximum loop delay. As an example, we assume that the fanout latch $j$ of the current latch $i$ is the starting latch. By adding the delay shift from $i$ to $j$, the cumulative delay shifts of the loops which are traversed can be computed. These cumulative delay shifts should be less than or equal to 0. Consider an item $D_{ji} + C_{ji}T$ in the arrival time $a_i$, we can write the loop constraint as

$$D_{ji} + C_{ji}T + m_{ij} \leq 0 \iff \quad (43)$$
$$D_{ji} + C_{ji}T + q_i + \Delta_{ij} - c_{ij}T \leq 0 \iff \quad (44)$$
$$(D_{ji} + q_i + \Delta_{ij})/(c_{ij} - C_{ji}) \leq T \quad (45)$$

where $c_{ij} - C_{ji}$ is positive.

Similar to $V_1$, we create a variable $V_3$ to represent the constraint that all loops are nonpositive. Each time when a constraint like (45) is created, we compute the maximum of $V_3$ and the variable on the left side of (45), and then update $V_3$ with the result. After the loop traversals from all latches are fulfilled, all loop constraints are merged to $V_3$. Therefore, the constraint $L_3$ can be represented by

$$V_3 \leq T \quad (46)$$

After the loop paths from a latch are traversed, we mark this latch as visited. This means the nonpositive constraint for all the loops through this latch has been specified. Therefore, we need not to propagate arrival times through the visited latches in following loop traversals. This can reduce the runtime of the loop constraint extraction remarkably.

### 3.6 Summary of timing model extraction for latch based circuits

As described by (23), we use the constraints $L_{R1}$ and $L_{R2}$ and $L_3$ to verify the timing of a latch based module. $L_{R1}$ and $L_3$ are specified in our method simply by (30) and (46), where $V_1$ and $V_3$ are known random variables computed from the original circuit during timing model extraction. To specify $L_{R2}$ for a primary input $k$, the constraint set $\mathcal{C}_k$ is used. Each item in $\mathcal{C}_k$ is in the form of (37). When verifying the timing performance of a hierarchical design, $\hat{a}_k$ is computed from the modules logically before the current module. As $\hat{a}_k$ becomes known, (37) can be rewritten as

$$(\hat{a}_k + D_{kj} + s_j)/(1 - C_{kj}) \leq T \quad (47)$$

As all the constraints can be written in the similar form in (30), (46) and (47), the constraints related to the current module in (23) can be easily represented by the maximum of all the random variables in (30), (46) and (47), where (47) is computed with the constraint items of all the primary inputs. Compared with directly verifying the timing performance of a latch based circuit, the timing constraints contained in an extracted timing model are very simple so that the statistical timing analysis of a hierarchical design can be accelerated drastically.

## 4. EXPERIMENTAL RESULTS

In this section, the results of the proposed method applied to the ISCAS89 benchmark circuits are shown. The algorithms were implemented in C++ and tested using a 2.33GHz CPU. The cells in the benchmark circuits were mapped to a 90nm library from an industry partner. The standard deviations of transistor length, oxide thickness and threshold voltage were assigned to 15.7%, 5.3% and 4.4% of the nominal values respectively [13]. The cell delays were created using the method proposed in [1]. We used the SSTA engine proposed in [18] to compute the addition and maximum of random variables, although our method is independent of the SSTA engine in use.

In order to verify the accuracy of the extracted timing models, we simulated the application context by generating arrival times for the primary inputs of each module randomly. For each circuit, we first computed the maximum mean $\mu_M$ and standard deviation $\sigma_M$ of the delays from all the primary inputs to their fanout sequential cells. We then generated $m$ random variables as the arrival times at the primary inputs, where $m$ is the number of primary inputs of the benchmark circuit except clock input. The means of these random variables were generated randomly between 0 and $0.5\mu_M$. The standard deviations were set between $0.05\sigma_M$ and $0.15\sigma_M$ randomly. The correlations between these random variables were set between 0.4 and 0.8, for the purposes of our experiments.

### 4.1 Results of timing model extraction for flip-flop based circuits

The results of applying the proposed timing model extraction method to the ISCAS89 benchmark circuits are listed in Table 1, where all sequential cells are assumed as flip-flops. Ac-cording to Section 2, the number of the variables contained in a timing model for a flip-flop based circuit is $m + n + 1$. $n$ is the number of primary outputs of the benchmark circuit. Compared with the number of combinational cells ($n_c$) and the number of sequential cells ($n_s$) in the benchmark circuits, we can see that the extracted timing models are much smaller than the original circuits.

Table 1: Results of timing model extraction for flip-flop based circuits

| Circuit | $m$ | $n$ | $n_c$ | $n_s$ | $\mu_{err}$ | $\sigma_{err}$ | $T_{err}^{97}$ | $t_{MC}$ | $t_M$ | $t_G$ |
|---|---|---|---|---|---|---|---|---|---|---|
| s298 | 3 | 6 | 119 | 14 | 0.06 | 0.55 | 0.18 | 5.17 | 0 | 0 |
| s526 | 3 | 6 | 193 | 21 | 0.06 | 0.42 | 0.15 | 9.34 | 0 | 0.01 |
| s820 | 18 | 19 | 289 | 5 | 0.01 | 0.28 | 0.05 | 18.82 | 0 | 0.01 |
| s1238 | 14 | 14 | 508 | 18 | 0.14 | 0.99 | 0.30 | 25.37 | 0 | 0.02 |
| s1423 | 17 | 5 | 657 | 74 | 0.14 | 0.31 | 0.05 | 45.51 | 0 | 0.02 |
| s5378 | 35 | 49 | 2779 | 179 | 0.32 | 0.27 | 0.19 | 246.41 | 0 | 0.13 |
| s9234 | 36 | 39 | 5597 | 211 | 0.65 | 0.43 | 0.42 | 733.12 | 0 | 0.3 |
| s13207 | 62 | 152 | 7951 | 638 | 0.36 | 0.91 | 0.14 | 981.93 | 0 | 0.48 |
| s15850 | 77 | 150 | 9772 | 534 | 0.68 | 0.41 | 0.64 | 1258.53 | 0 | 0.93 |
| s38584 | 38 | 304 | 19253 | 1246 | 0.36 | 0.29 | 0.26 | 3403.98 | 0.01 | 1.19 |
| average | | | | | 0.28 | 0.49 | 0.24 | | | |

Table 2: Results of timing model extraction for latch based circuits

| Circuit | $n_i^c$ | $n_o^c$ | $\mu_{err}$ | $\sigma_{err}$ | $T_{err}^{97}$ | $t_{MC}$ | $t_M$ | $t_G$ |
|---|---|---|---|---|---|---|---|---|
| s298 | 2 | 5 | 0.09 | 0.05 | 0.06 | 21.27 | 0 | 0.01 |
| s526 | 4 | 11 | 0.10 | 1.08 | 0.33 | 42.87 | 0 | 0.01 |
| s820 | 2 | 36.8 | 0.04 | 0.69 | 0.10 | 44.18 | 0 | 0.03 |
| s1238 | 1 | 13.9 | 0.14 | 0.99 | 0.30 | 95.28 | 0 | 0 |
| s1423 | 4.4 | 31.4 | 0.19 | 1.04 | 0.37 | 556.4 | 0 | 1.94 |
| s5378 | 2.2 | 55.7 | 0.89 | 1.31 | 0.98 | 2445.7 | 0 | 2.73 |
| s9234 | 7.4 | 81.3 | 0.17 | 0.58 | 0.23 | 3578.1 | 0 | 18.32 |
| s13207 | 1.3 | 3.7 | 0.30 | 0.47 | 0.16 | 5031.7 | 0.01 | 8.58 |
| s15850 | 2.7 | 31.6 | 0.48 | 0.20 | 0.44 | 21439.4 | 0 | 174.08 |
| s38584 | 2.4 | 6.8 | 0.62 | 0.24 | 0.56 | 54610.5 | 0 | 307.32 |
| average | | | 0.30 | 0.67 | 0.35 | | | |

To verify the accuracy of the extracted timing models, we compare the results of timing analysis using the extracted timing models and the results of Monte Carlo simulation with 10000 iterations. In table 1, $\mu_{err}$ is defined as $(|\mu_{model} - \mu_{MC}|/\mu_{MC}) \times 100$, i.e. in percentage, where $\mu_{model}$ and $\mu_{MC}$ are the means of the clock period using the timing models and from Monte Carlo simulation using the original circuits respectively. Similar to $\mu_{err}$, $\sigma_{err}$ is defined to show the accuracy of standard deviation of the clock period. From Table 1, we can see that the extracted timing models are very accurate. Additionally, for the clock period at which 97% yield can be achieved, the error comparing our proposed method to Monte Carlo simulation is shown as $T_{err}^{97}$, also in percentage. This confirms the accuracy of the extracted timing models as well.

### 4.2 Results of timing model extraction for latch based circuits

To verify the quality of the timing models for latch based circuits, we assume all the sequential cells in the ISCAS89 benchmark circuits are latches. For experiment, we test these circuits with only one clock phase and all the enabling clock edges are set to 0.5 times the clock period. The predefined threshold $\delta$ for the probability comparison with (34) and (42) is 99.9%, which is very close to 1 so that the removal of arrival time items during setup time constraint extraction affects the accuracy of timing models only with very small probability.

To verify the accuracy of these timing models, we also run Monte Carlo simulation with 10000 iterations for each benchmark circuit. In each iteration, the method proposed in [15] is used to compute the clock period of the original benchmark circuit. The results are shown in Table 2, where the notations have the same definitions as for flip-flop based circuits. The average number of constraints for each primary input is shown as $n_i^c$. The average number of delay items to primary outputs is shown as $n_o^c$. Because of latch transparency, arrival times from many internal latches and from primary inputs can reach primary outputs. Consequently, $n_o^c$ is relatively larger than the number of the constraints for primary inputs. From $\mu_{err}$, $\sigma_{err}$ and $T_{err}^{97}$ we can draw the conclusion that the extracted timing models are very accurate.

### 4.3 Runtime comparison

The runtimes of the Monte Carlo simulation and timing analysis using the extracted timing models are shown as $t_{MC}$ and $t_M$ in seconds in both tables. The 0s for $t_M$ mean that the runtimes are shorter than $10^{-6}$s, thus can not be measured accurately using the clock function. The runtimes to extract the timing models, shown as $t_G$ in seconds, are roughly equal to the runtimes of statistical timing analysis of the original circuits using previous approaches ([18] for flip-flop based circuits, and [3, 21] for latch based circuits). Comparing $t_G$ and $t_M$, we can see that timing analysis using our proposed timing models gains many orders of magnitude in runtime acceleration.

## 5. CONCLUSION

In this paper, we proposed a method to extract timing models of sequential circuits for statistical timing analysis effectively. Compared with the original circuits, the extracted timing models are very small and the runtime of timing analysis can be drastically reduced when such timing models are used. From our experiments, the mean and standard deviation of the clock period have less than 1% error on average, when comparing the results using our timing models with Monte Carlo simulation.